\documentclass[reprint,amsmath,amssymb,aps,prb,superscriptaddress,floatfix,nofootinbib,nobibnotes]{revtex4-2}

\pdfoutput=1
\usepackage{graphicx}
\usepackage{dcolumn}
\usepackage{bm}
\usepackage{braket}
\usepackage[mathlines]{lineno}
\usepackage{stackengine}
\usepackage{verbatim}
\usepackage{graphics}
\usepackage[colorlinks]{hyperref}
\hypersetup{linkcolor = blue, citecolor = blue}
\usepackage{xcolor}
\usepackage{appendix}
\usepackage{soul}
\usepackage{ulem} 
\usepackage{multirow}
\usepackage{amsmath}

\begin{document}
\preprint{APS/123-QED}
\title{Escaping AB caging via Floquet engineering: photo-induced long-range interference in an all-band-flat model}
\author{Aamna Ahmed}
\email{aamna.ahmed203@gmail.com}
\affiliation{Institute of Physics, University of Augsburg, Germany}
\author{M\'onica Benito}%
\email{monica.benito@uni-a.de}
\affiliation{Institute of Physics, University of Augsburg, Germany}
\affiliation{Center for Advanced Analytics and Predictive Sciences, University of Augsburg, Germany}
\author{Beatriz P\'erez-Gonz\'alez}%
\email{perezg.bea@gmail.com}
\affiliation{Institute of Physics, University of Augsburg, Germany}

\date{\today}

\begin{abstract}
Flat-band lattices hosting compact localized states are highly sensitive to external modulation, and the tailored design of a perturbation to imprint specific features becomes relevant. Here we show that periodic driving in the high-frequency regime transforms the all-flat-band diamond chain into one featuring two tunable quasi-flat bands and a residual flat band pinned at $E=0$. The interplay between lattice geometry and the symmetries of the driven system gives rise to drive-induced tunneling processes that redefine the interference conditions and open a controllable route to escaping Aharonov–Bohm caging. Under driving, the diamond chain effectively acquires the geometry of a dimerized lattice, exhibiting charge oscillations between opposite boundaries. This feature can be exploited to generate two-particle entanglement that is directly accessible experimentally. The resulting drive-engineered quasi-flat bands thus provide a versatile platform for manipulating quantum correlations, revealing a direct link between spectral fine structure and dynamical entanglement.
\end{abstract}

\maketitle


\section{INTRODUCTION}\label{sec:level1}

Flat-band systems~\cite{PhysRevLett.81.5888, PhysRevLett.85.3906, Leykam01012018, Aoki21102025} have attracted wide interest due to their lattice geometries that enforce exact destructive interference, producing dispersionless bands with compact localized states (CLS) \cite{PhysRevLett.121.075502} confined to a small region of sites. Unlike Anderson-localized states \cite{PhysRev.109.1492}, which originate from disorder, CLS emerge from perfect interference between distinct hopping paths, which prevents particles from leaking out. 

This flat-band condition is fragile, making these systems exquisitely sensitive to perturbations such as modified hoppings or disorder~\cite{PhysRevB.98.235109, PhysRevB.88.224203, PhysRevB.106.205119, PhysRevB.107.245110, Marques2024}. However, this sensitivity is part of the appeal: flat-band systems provide a tunable platform where different phases can be realized by introducing tailored perturbations or additional interactions, including correlated phases~\cite{doi:10.1142/S0217979215300078}, such as ferromagnetism~\cite{PhysRevLett.121.096401,10.1143/PTP.99.489,Park2025}, superconductivity~\cite{PhysRevLett.121.087001,Tian2023}, topological bands and fractional Chern insulators~\cite{doi:10.1142/S021797921330017X, PhysRevLett.109.186805, PhysRevLett.125.266403, PhysRevA.99.023612}. Recent experiments have realized these geometries in ultracold atoms, photonic lattices, and superconducting circuits implementing Lieb, Kagome, and rhombic configurations~\cite{PhysRevLett.114.245504, Real2017, TangSongXiaXiaMaYanHuXuLeykamChen+2020+1161+1176, doi:10.1126/sciadv.adj7195, Song2025, Yang2024, Chen2025, lape2025realizationcharacterizationallbandsflatelectronic, Kollár2019, PhysRevResearch.2.043426, PhysRevLett.121.075502, Kim2025, bt9s-qsfj, Li2023, Drost2017, doi:10.1126/sciadv.aau4511, Mukherjee15}. 

A specially fine-tuned regime is that of Aharonov–Bohm (AB) caging \cite{PhysRevLett.129.220403, PhysRevLett.81.5888, Longhi14, PhysRevB.64.155306}, which yields an all-band-flat (ABF) system as a result of complete destructive interference among the eigenstates induced by a magnetic flux threading each plaquette in the lattice. In this condition, every eigenstate forms a compact localized state (CLS), with no coexistence of extended Bloch modes.

Periodic driving, on the other hand, is a widely used tool for engineering phases of quantum materials \cite{floquetmaterials, Bloch2022}. The idea of using time-periodic modulations of lattice parameters or fields, such as AC electric fields, lattice shaking, or microwave modulations, as an external knob for reshaping band structures is commonly known as Floquet engineering~\cite{Bukov04032015, PhysRevX.4.031027, GRIFONI1998229}, with many celebrated examples, ranging from controllable band structure engineering~\cite{GRIFONI1998229, Zhou2023, Bao2022} and topological properties~\cite{PhysRevB.79.081406, PhysRevB.82.235114, PhysRevLett.123.126401, PhysRevB.90.205127, Rudner2020, PhysRevResearch.1.023031, Merboldt2025, doi:10.1126/science.1239834, PhysRevB.88.245422, PhysRevLett.112.156801}, to time crystals~\cite{Anisur2025-kh, PhysRevLett.117.090402, PhysRevLett.118.030401, Zhang2017, PhysRevB.106.L060305}, dynamical freezing~\cite{PhysRevB.91.121106,PhysRevX.11.021008}, dynamical phase transitions~\cite{Hamazaki2021}, and driven phases without static counterpart \cite{PhysRevB.99.195133,  PhysRevB.109.064303}.

Over the past decade, considerable effort has been devoted to understanding how Floquet modulation can be employed to fine-tune flatband structures, alter their localization properties, or induce new transport or topological phenomena~\cite{PhysRevB.104.134307, PhysRevLett.105.086804, Song2025, PhysRevB.105.245136, PhysRevResearch.2.043275, 10.1088/1367-2630/adfd07, Song2025, kumar2025bandstructureevolutionkagome, Maczewsky2017, PhysRevB.108.045415, Kang2020}, granting dynamic control over the single-particle spectrum. A complementary question, however, is whether deliberately destroying the flat-band character through a Floquet drive~\cite{PhysRevLett.116.245301, Li2025} can actually be advantageous to enhance the functionality of envisioned quantum networks based on these geometries \cite{PhysRevLett.123.080504, Taie2020, Zurita2023fastquantumtransfer, 10.1063/5.0153770} while retaining or transforming key features inherited from the undriven system. 

In this work, we investigate the ABF diamond lattice under a periodic on-site square-wave drive. In the undriven case, with a flux of $\pi$ per plaquette, the lattice exhibits perfectly flat bands. Introducing a high-frequency drive, we explore the interplay between lattice geometry and temporal modulation, and how the interference conditions are dynamically redefined. Although the exact ABF condition is lifted, yielding two quasi-flat bands and one flat band, the drive induces a new form of long-range connectivity within the lattice that enables controlled escape from Aharonov–Bohm caging, based on photo-induced long-range hoppings. Building upon this mechanism, we propose a protocol for dynamically generating entanglement between particles across a bipartition: at intermediate driving frequencies, the quasi-flat bands give rise to coherent oscillations between separable and entangled states.

The paper is organized as follows. Section~\ref{sec:model} introduces the model and driving protocol, outlining the key features of the $\pi$-flux diamond chain. Section~\ref{sec:highfreqex} presents analytical results characterizing the modified interference conditions through a high-frequency expansion and the derivation of an effective, time-independent Hamiltonian. Section~\ref{sec:dynamics} illustrates these concepts via charge dynamics in the bulk of the driven system. Section~\ref{sec:entanglement} extends the analysis to a finite interacting system, where we propose and analyze a protocol for generating and stabilizing entangled states, tracing the origin of oscillations in the entanglement measures. Finally, Section~\ref{sec:conclusions} summarizes the main conclusions.

\section{MODEL}\label{sec:model}

\begin{figure}
    \centering
\includegraphics{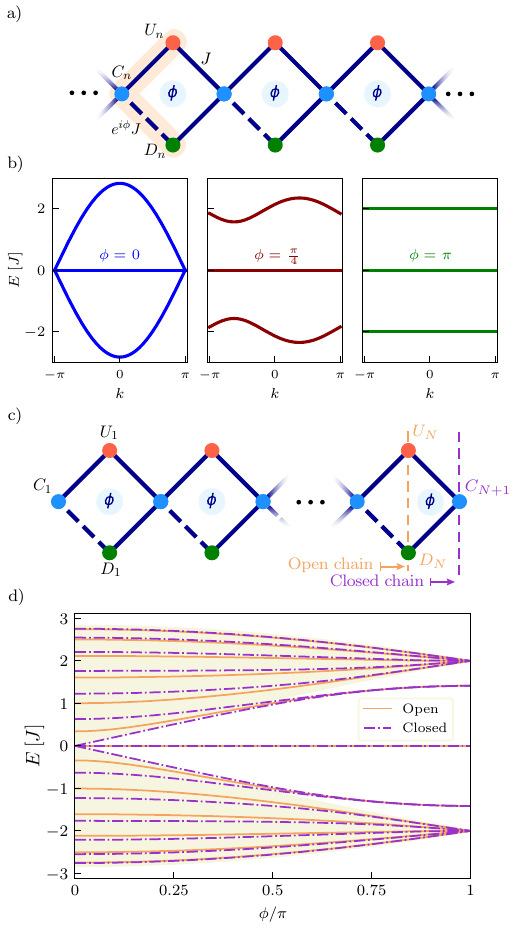}
    \caption{\textbf{a)} Schematic representation of the diamond lattice with a $\phi$ flux threading each plaquette. The unit cell, containing $\{C_n, U_n, D_n\}$ sites (depicted in blue, red, and green, respectively), is shaded in orange. All hoppings have magnitude $J$, while the one connecting $C_n \leftrightarrow D_n$ has an extra Peierls phase accounting for the net flux. \textbf{b)} Band structure of the system for different flux choices. Only $\phi = \pi$ corresponds to an all-flat-band system. \textbf{c)} Schematic representation of a finite chain with $N$ cells and two possible terminations: open end (with $\{U_N, D_N\}$ as the boundary), and closed end ($C_{N+1}$ as boundary). \textbf{d)} Energy spectrum for a finite-size chain of $N = 6$ cells, and $J = 1$, for both open (solid orange) and closed (dashed purple) end. The bulk is shaded in yellow. The main difference relies on the presence of two pairs of edge states in each gap for the closed chain.}
    \label{fig:schematic}
\end{figure}

\paragraph*{\textbf{Static part: bulk Hamiltonian and finite-size chains.}} The $\phi$-flux diamond chain consists of a three-site unit cell connected by nearest-neighbor hoppings of amplitude $J$, as shown in the schematic of Fig. \ref{fig:schematic} a). Its Bloch Hamiltonian can be written as $\hat{H}_{0} = \sum_k \hat{\Psi}_k^\dagger \, \hat{H}_{0} (k) \, \hat{\Psi}_k$, with
\begin{equation}
    \hat{H}_{0} (k)  = J \left( \begin{array}{ccc}
    0 & 1+e^{-ik} & e^{-i\phi} + e^{-ik}\\
    1+e^{ik} & 0 & 0\\
    e^{i\phi} + e^{ik} & 0 & 0 
    \end{array} \right)
    \label{eq:hamiltonian_or}\ ,
\end{equation}
where $k$ denotes crystal momentum and $\phi$ represents a net magnetic flux threading each of the plaquettes. The corresponding spinor $\hat{\Psi}_k = (\hat{C}_k,\hat{U}_k,\hat{D}_k)$ is defined as the weights of the Bloch wavefunctions on the central, upper and lower sites, respectively.

Tuning the value of $\phi$ has a crucial impact on the band structure: for $\phi=0$, the system hosts a single flat band at $E_0=0$ and two dispersive bands $E_{\pm}(k)=\pm 2\sqrt{2}J\cos(k/2)$, forming a gapless spectrum, whereas a finite flux opens two gaps and reduces the curvature of $E_{\pm}(k)$. At $\phi=\pm\pi$, destructive interference is maximized and all three bands become completely flat, with $E_{\pm}=\pm 2J$: all eigenstates are CLS. The resulting band structure for different choices of $\phi$ is shown in Fig. \ref{fig:schematic}b). In the following, we will fix $\phi = \pi$, as we are mainly interested in the all-band-flat (ABF) model, and $J = 1$.

In finite chains, the bulk spectrum described above is complemented by \textit{boundary modes}. In fact, the $\pi$-flux diamond chain provides a striking example of a \textit{square-root topological insulator}, featuring in-gap states pinned at $E_{\mathrm{e.s.}}=\pm \sqrt{2}J$ that inherit topological protection from a parent Hamiltonian \cite{Kremer2020, Zurita2021tunablezeromodes, ZuritaHiddenTop}. Unlike the exponentially localized edge states of conventional one-dimensional topological chains, such as the Su–Schrieffer–Heeger (SSH) model \cite{AsbothOroszlanyPalyi2016, PhysRevLett.42.1698}, these modes are compactly localized at the $C$-site terminations, vanishing exactly outside the finite region of the corresponding AB cage.

In this work we consider finite chains of $N$ unit cells plus an additional $C$ termination, as shown in Fig. \ref{fig:schematic}c). With this choice, all plaquettes are closed, resulting in a total of $3N +1$ sites. The Hamiltonian is given by 
\begin{eqnarray}
    \hat{H}_{\mathrm{finite}} & = & J\sum_{n=1}^{N} \left( \hat{C}^\dagger_n\, \hat{U}_n + e^{i\phi} \hat{D}_n^\dagger\, \hat{C}_n + \hat{C}_{n+1}^\dagger \, \hat{D}_n \right. \nonumber \\
    && \quad \quad\quad\quad\quad\quad\quad\quad \left. + \hat{U}^\dagger_n\, \hat{C}_{n+1} + \text{h.c.}\right) \ .
\end{eqnarray}
In an open chain without an additional $C$ site, $\hat{C}_{N+1}^\dagger \, \hat{D}_N + \hat{U}^\dagger_N\, \hat{C}_{N+1} + \text{h.c.} = 0$. Fig. \ref{fig:schematic}d)  compares the band structure of $\hat{H}_0(k)$ to the energy states obtained from a finite chain with $N =6$ unit cells, with both a closed (extra $C$ site, $3N+1$ sites) and open end ($3N$ sites), as a function of $\phi$. In the configuration with an extra $C$ site, the spectrum reorganizes to host four edge states, appearing in two pairs. Each end of the chain terminates in a $C$, so both of them host boundary modes (see Appendix \ref{app:edgestates}). As it becomes clear by simple inspection of the energy spectrum and its symmetry about the $E = 0$ axis, the diamond chain with this hopping configuration is \textit{bipartite}: the $C$ ($U/D$) sites form the minority (majority) sublattice. In the site basis this bipartition ensures the presence of \textit{chiral symmetry operator} \cite{PhysRevB.96.161104, Călugăru2022}, represented by the operator $\hat{\Gamma} = \mathrm{diag}(+1,-1,-1,+1,-1,-1,\dots)$,  that fulfills $\hat{H}_{\mathrm{finite}} = - \hat{\Gamma} \hat{H}_{\mathrm{finite}} \hat{\Gamma}^{-1}$. 

Taken together, these features make the diamond chain a paradigmatic setting to investigate the interplay of flat bands and caging under external perturbations. The $\pi$-flux configuration is particularly fine-tuned: even small deviations of the flux restore dispersive character to the bands. A central question, therefore, is to what extent the distinctive features of the chain survive in the presence of additional driving fields, and whether such features can be exploited for control and transport.\\
 
\paragraph*{\textbf{Driving protocol and Floquet theory.}} In this work we consider an additional time-dependent field, comprising a square-wave modulation of the onsite energies of the $U$ and $D$ sites of the chain,
\begin{equation}
    \hat{H}_\text{d}(k,t) = f(t) \sum_{k}^{N }\left( \hat{U}_k^\dagger \hat{U}_k + \zeta \, \hat{D}^\dagger_k \hat{D}_k\right),
\end{equation}
with 
\begin{equation}
    f(t) =
    \begin{cases}
    +A, & 0 \leq t < \frac{T}{2}, \\
    -A, & \frac{T}{2} \leq t < T ,
    \label{eq2}
    \end{cases}
\end{equation}
where $T = 2\pi/\omega$ is the driving period, $\omega$ the driving frequency, and $A$ the driving amplitude. The extra factor $\zeta$ lets us introduce a difference between both modulations in the $U/D$ sites. 
We consider two cases: $\zeta = 1$ and $-1$, corresponding to an in-phase (symmetric) / out-of-phase (antisymmetric) driving. For a finite chain, the driving term would be diagonal in the site basis as well, with $\hat{H}_\text{d}(t) =f(t) \sum_{n = 1}^{N}\left( \hat{U}_n^\dagger \hat{U}_n + \zeta \, \hat{D}^\dagger_n \hat{D}_n\right)$. 

In general, the presence of an on-site, static contribution in the Hamiltonian breaks chiral symmetry, which implies that the eigenstates will not come in chiral pairs $E \leftrightarrow -E$ anymore. However, there is a key distinction between the symmetric/antisymmetric case \cite{PhysRevB.106.205119}. Caging does survive to equal $U/D$ onsite energies, since $C$ sites still couple identically to $U/D$: all bands remain flat, but shifted in energy according to the value of the perturbation (see Appendix \ref{app:onsite_energies_static}). 

In our case, the on-site term is modulated in time, and the total Hamiltonian of the system yields
\begin{equation}
\hat{H}_{\mathrm{tot}}(k, t) = \hat{H}_{0}(k) + \hat{H}_{\mathrm{d}}(k, t).
\label{eq:total_hamiltonian}
\end{equation}
To investigate how a time-periodic on-site modulation with period $T$ modifies the properties of the static system, we employ Floquet theory. Solutions to the time-dependent Schrödinger equation take the form $\ket{\psi_n(t)} = e^{-i\varepsilon_n t}\ket{u_n(t)}$, where $\varepsilon_n$ denotes the quasienergy and $\ket{u_n(t)}$ is the corresponding  Floquet mode. The quasienergies play the role of energies in static systems, defining a quasienergy spectrum, while the Floquet modes inherit the time-periodicity of the Hamiltonian, $\ket{u_n(t)}=\ket{u_n(t+T)}$. Importantly, the quasienergy spectrum consists of infinitely many replicas shifted by integer multiples of the driving frequency $\omega$, since a shift $\varepsilon_n \to \varepsilon_n+m\omega$ corresponds to the same physical state $\ket{\psi_n(t)}$.

In the high-frequency regime, the quasienergy replicas are well separated, allowing the dynamics of the driven system to be described within a single replica. In this case, systematic high-frequency expansions yield an effective time-independent Hamiltonian $\hat{H}_{\mathrm{eff}}$ \cite{Eckardt_2015, Bukov04032015, PhysRevX.4.031027} that governs the stroboscopic evolution and can be analyzed using standard techniques for static systems. The primary effect of the periodic drive is to renormalize and reshape the band structure, thereby enabling external control of the system’s properties through the driving parameters. In the following section, we adopt this approach to derive an effective Hamiltonian for the driven diamond chain, which reproduces the essential features of the quasienergy spectrum and allows us to analyze the impact of the drive on the symmetries of the lattice, Aharonov–Bohm caging, flat bands, and edge states. \\

\begin{figure}
    \centering
    \includegraphics{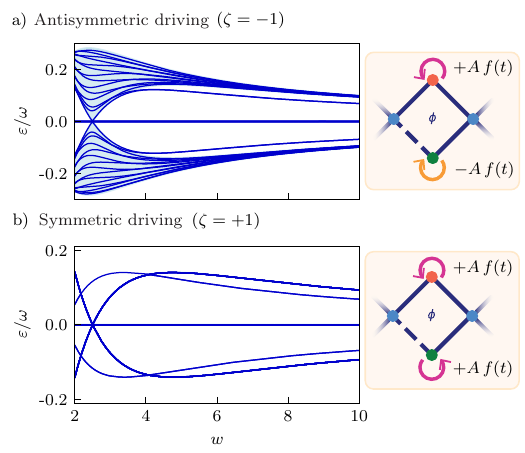}
    \caption{Quasienergy spectrum $\varepsilon / \omega$ as a function of the driving frequency, for a finite-diamond lattice of $N = 14$ cells, $J = 1$ and $\phi = \pi$, with a closed end, under \textbf{a)} antisymmetric ($\zeta = -1$) and \textbf{b)} symmetric  ($\zeta = 1$) driving of the onsite potentials. The bulk is shaded in blue. The amplitude is fixed so that $\tilde{A} = A/ \omega = 1.2$} \label{fig:quasienergies}
\end{figure}

\paragraph*{\textbf{Quasienergy spectrum.}} We first compute the quasienergy spectrum of the driven diamond chain numerically as a function of the driving frequency for $\omega> 2J$ (Fig.~\ref{fig:quasienergies}). The spectrum retains chiral symmetry for both $\zeta=\pm 1$, featuring a flat band pinned at $E_0=0$ and two symmetric branches with $E_{+}=-E_{-}$. The edge states remain degenerate, consistent with the preservation of the main symmetries. Interestingly, for $\zeta=-1$ the upper and lower bands develop finite dispersion, which increases as the driving frequency is reduced. By contrast, for $\zeta=1$ all three bands remain perfectly flat.

For comparison, recall that in the static chain a flux $\phi\neq \pi$ through each plaquette also induces dispersion in the upper and lower bands. In the driven case, however, a more careful analysis is required to identify the physical mechanism responsible for the observed broadening, and in particular to clarify the qualitative difference between symmetric and antisymmetric driving. 

Additionally, static on-site contributions, either symmetric or antisymmetric, break chiral symmetry, whereas time-modulated ones seem to render a chiral spectrum. Even though in the instantaneous Hamiltonian chiral symmetry is broken, ensuring its presence or absence in the quasienergy spectrum is connected to the symmetry properties of the driving field \cite{PhysRevB.90.125143, PhysRevB.109.184307}, and to finding a chiral operator $\hat{\Gamma}$ fulfilling $\hat{\Gamma}^\dagger \hat{H}_{\mathrm{tot}}(k,t) \hat{\Gamma} = - \hat{H}_{\mathrm{tot}}(k,-t)$. In our case, the square-wave time modulation meets the requirement by splitting the driving period into two parts naturally. 

\section{High-frequency expansion and analytical result \label{sec:highfreqex}}

\paragraph*{\textbf{Effective Hamiltonian.}} To understand  these features, we first rotate the Hamiltonian in Eq. \eqref{eq:total_hamiltonian} into the frame defined by $\hat{U}(t) = \mathrm{exp}\left[ -i \int \text{d}t\,\hat{H}_{\mathrm{driv}}(t)\right]$, to encode the effect of strong driving fields, 
\begin{eqnarray}
\hat{H}^{\prime}_{\mathrm{tot}} (k, t) & = & J \left( \begin{array}{ccc}
    0 & (*) & (*) \\
    e^{iF(t)}(1+e^{ik}) & 0 & 0\\
    e^{i\zeta F(t)}(e^{i\phi} + e^{ik}) & 0 & 0 
    \end{array} \right)
    \label{eq:transformed_H} \ ,
\end{eqnarray}
where $F(t) = \int f(t)$. The notation $(*)$ indicates complex conjugate of the matrix element with the transposed indexes, as corresponds to an Hermitian operator. Then, we use Van Vleck perturbation theory to obtain a high-frequency expansion of the driven Hamiltonian, defined in terms of inverse powers of $\omega$, $\hat{H}_{\mathrm{eff}} = \sum_{n = 0}^{\infty} (1/\omega^n) \, \hat{H}_{\mathrm{eff}}^{(n)}$ (see Appendix \ref{app:effective_hamiltonian} for details). 

If the frequency is high enough, the zeroth order of this expansion,
\begin{equation}
    \hat{H}_{\mathrm{eff}}^{(0)}(k) = \frac{1}{T}\int_0^T \text{d}t\,\hat{H}^\prime_{
    \mathrm{tot}}(k,t) \ ,
    \label{eq:effective_h}
\end{equation}

is typically enough to reproduce the quasienergy spectrum of the driven system.
In this case, the net effect of the driving is the renormalization of the hopping amplitudes, through \cite{PhysRevLett.123.126401}
%
%
%
\begin{eqnarray}
    J & \rightarrow & \,J_0\left(\tilde{ A}\right) = \frac{i J}{\tilde{A} \pi}\left[ 1 - \mathrm{exp}\left(i \tilde{A} \pi \right)\right],
    \label{eq:effective_J0}
\end{eqnarray}
where we have defined $\tilde{A} = A/\omega$. For antisymmetric driving, the hopping $C \rightarrow D$ takes into account the extra phase $\zeta = -1$ in $\mathrm{exp}(i\zeta F(t))$, so that $J_0(-\tilde{ A}) = J_0^*(\tilde{ A})$, and the effective Hamiltonian reads
%
%
\begin{equation}
    \hat{H}_{\mathrm{eff}}^{(0)}(k, \zeta = -1) = \left( 
    \begin{array}{ccc}
    0 & (*)& (*) \\
    J_0(1+e^{ik}) & 0 & 0 \\
    J_0^*(-1+e^{ik})& 0 & 0
    \end{array} \right) \ .
    \label{eq:h_eff0}
\end{equation}
For $\zeta = +1$, both $C \rightarrow U$ and $C \rightarrow D$ hoppings are renormalized with $J_0(\tilde{ A})$. The results of the calculations for symmetric drive are shown in Appendix \ref{sec:symmetric}.

Note that the net flux through each plaquette remains unchanged at its original value of $\phi = \pi$, although the photo-dressed hoppings acquire complex phases. This hopping renormalization can explain the new dependence of the band structure on the driving parameters, as expected. However, it fails to explain the difference between the symmetric and antisymmetric driving protocol, and in particular, the essential mechanism induced by the drive for $\zeta = -1$, which is the broadening of the upper and lower bands. AB caging relies on exact phase cancellation, so small photon-assisted corrections can restore dispersion even for large $\omega$. In this scenario, higher-order processes involving virtual photon exchange become the dominant source of new features.

Let us first analyze the case of antisymmetric driving, $\zeta = -1$. The second-order effective Hamiltonian has the  structure
\begin{equation}
\hat{H}_{\mathrm{eff}}^{(2)}(k, \zeta = -1) =\left( \begin{array}{ccc}
0 & (*) & (*) \\
 \mathcal{C} \,  h_{cu}(k) &  0 & 0 \\
\mathcal{C}^* \, h_{cd}(k) & 0 & 0
 \end{array}
\right) \ , \label{eq:Heff2}
\end{equation}
with a common factor containing the system and driving parameters,
\begin{equation}
    \mathcal{C} = \frac{2 \Gamma}{3\pi^3} \,\frac{J^3 \tilde{A}}{\omega^2}\sin\left(\pi \tilde{A} \right)\left[1 + e^{ i \tilde{A} \pi }\right] \ ,
\end{equation}
which includes the dimensionless coefficient $\Gamma$ containing all the $(m,n)$-harmonic-dependence,
\begin{equation}
    \Gamma =  \sum_{\substack{m \in\text{odd}\, \mathbb{Z}\\ n\in \mathbb{Z} \\ m,n \neq 0, m\neq n}}\frac{1}{m^2} \left[ \frac{2\tilde{A}^2(-1)^n m}{n \, \Pi_n \, \Pi_m \Pi_{n-m}} - \frac{3}{\Pi_m^2} \right]
    \ , \label{eq:gamma_Def}
\end{equation}
with the pole structure defined as $\Pi_{\alpha} = \left(\tilde{A}^2 - \alpha^2\right)$ (the convergence properties of the series coefficient $\Gamma$ are analyzed in Appendix \ref{sec:convergence}). Lastly, we 
also define the dispersive terms in Eq.~\eqref{eq:Heff2},
\begin{eqnarray}
    h_{cu} (k) & = & 5(1 + e^{ik}) +3(e^{-ik} + e^{2ik}), \label{eq:hoppings_ctou}\\
    h_{cd} (k) & = & 5(- 1 +  e^{ik}) +3( e^{-ik}  - e^{2ik}) \label{eq:hoppings_ctod}.
\end{eqnarray}

\paragraph*{\textbf{Long-range hoppings.}} Let us now analyze the structure of the effective Hamiltonian. The most relevant feature lies in the new hopping structure induced by the driving, which becomes explicit in the second-order effective Hamiltonian, Eq.~\eqref{eq:Heff2}, and reveals the coexistence of two hopping networks within the same lattice: a primary (nearest-neighbor) and a secondary (next-nearest-neighbor) one. 

\begin{figure}
    \centering
    \includegraphics{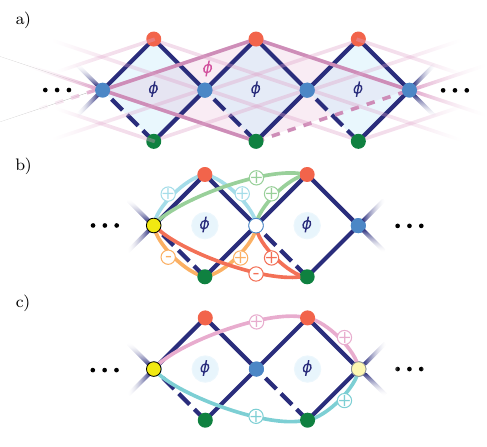}
    \caption{Schematic of the diamond lattice and its connectivity in the high frequency regime for an antisymmetric driving. \textbf{a)}  Together with the pre existing hoppings (blue), the  photo-induced second-neighbor hoppings are represented dark pink. The dashed lines represent negative hopping. \textbf{b)}, \textbf{c)} The curved lines represent the paths that connect one central site $C_n$ (depicted in yellow) with $C_{n+1}$ and $C_{n+2}$. The signs indicate the phase picked up during each hopping process. With this, we can understand the destructive interference happening for the process in b).}
    \label{fig:secondneig}
\end{figure}

First, the nearest-neighbor hoppings already contained in $\hat{H}_{\mathrm{eff}}^{(0)}$ acquire additional contributions $5(\pm 1 + e^{ik})$, which are suppressed by $1/\omega^2$ and therefore constitute small corrections to the leading-order amplitude $J_{0}$. More importantly, the drive generates a secondary diamond lattice formed by second-neighbor hoppings, proportional to $3(e^{-ik}\pm e^{2ik})$.
Remarkably, the $\pi$-flux condition is preserved not only across the original plaquettes but also through the enlarged plaquettes (extending over three plaquettes in the original lattice) of the emergent lattice, as reflected in the $e^{i\pi}=-1$ phase factors in Eqs. \eqref{eq:hoppings_ctou} and \eqref{eq:hoppings_ctod}.
Thus, \textit{the drive maintains the original $\pi$-flux configuration while imprinting it identically onto the newly formed secondary lattice}. This is depicted in Fig. \ref{fig:secondneig}a. The new effective unit cell in the driven chain is, naturally, enlarged, due to the presence of long-range hoppings.

The coexistence of hopping networks breaks the stringent conditions required for perfect Aharonov–Bohm (AB) caging, as the additional hopping paths modify the interference pattern. In the quasienergy spectrum corresponding to the antisymmetric driving scheme (Fig.~\ref{fig:quasienergies}a), this manifests as a broadening of the upper and lower bands, which acquire finite dispersion, even though the net flux remains fixed at $\phi=\pi$ for both hoppings structures.

For completeness, let us now comment on the effective Hamiltonian under symmetric driving $\hat{H}^{(2)}_{\mathrm{eff}}(\zeta = +1) $. While its structure resembles that of $\hat{H}^{(2)}_{\mathrm{eff}}(k,\zeta = -1)$ (see Appendix \ref{sec:symmetric} for the complete expression), it is particularly relevant to look into its dispersive part,  
\begin{equation}
    h_{cu}^{(\text{S})}(k) = 1 + e^{ik}, \hspace{10pt} h_{du}^{(\text{S})}(k) = -1 + e^{ik}.
    \label{eq:dispersive_part}
\end{equation}
Interestingly, the effective Hamiltonian reveals that the symmetric driving does not induce any long-range hoppings in the lattice. The second-order term of the high-frequency expansion merely renormalizes the existing nearest-neighbor couplings, without generating new connections.

This contrast highlights a subtle interplay between the driving protocol and the lattice geometry. The antisymmetric modulation breaks the mirror symmetry between the two arms of the $\pi$-flux diamond chain, dynamically perturbing the destructive interference in the static system. In this case, the modulation of the hopping amplitudes connecting $C_n \leftrightarrow U_n$ and $C_n \leftrightarrow D_n$ causes the interference to oscillate periodically between constructive and destructive. Over many driving cycles, this time-dependent imbalance gives rise to effective virtual pathways that extend beyond a single plaquette, coupling next-nearest cells.
In contrast, the symmetric modulation preserves the mirror symmetry and thus the interference condition responsible for localization. Under the gauge transformation $\hat{U}(t)$ that removes the on-site modulation, both arms acquire identical time-dependent phases, leaving their relative phase, and consequently the destructive interference, unchanged. As a result, the Floquet expansion yields only a renormalization of the nearest-neighbor couplings, without introducing new long-range terms or hybridizing compact localized states across plaquettes.

The presence of a flat band pinned at $E_0 = 0$ in the driven system and for both $\zeta = \pm 1$ can also be explained through a similar reasoning. First, recall that also for $\phi \neq \pi$, the central band of the rhombic lattice remains entirely flat, which means that its presence does not require the fine-tuning of the perfect destructive interference between all paths connecting $C_n \leftrightarrow C_{n+1}$. The CLS belonging to the central band are non-overlapping and therefore, perturbations that are uniform across all cells, like the driving protocol discussed in this work, will not couple them \cite{PhysRevB.95.115135, Flach_2014}. 

Analyzing the low-frequency part of the plot in Fig. \ref{fig:quasienergies}, we can expect that further orders of the effective Hamiltonian will not modify this reasoning, and therefore all the features of the driven system are qualitatively captured by the effective Hamiltonian obtained to second order.

\section{Interference and single-particle dynamics: decorated SSH chains \label{sec:dynamics}}

\paragraph*{\textbf{Dynamical selection rules.}} From the symmetry of the driven lattice and the structure of the generated hoppings, the qualitative impact of the antisymmetric drive can be anticipated.
Consider, for instance, two adjacent $C$ sites, $C_n$ and $C_{n+1}$.
In the undriven chain, a particle propagating between them can only follow two nearest-neighbor paths, $C_n \rightarrow U_n \rightarrow C_{n+1}$ and $C_n \rightarrow D_n \rightarrow C_{n+1}$, whose relative phase of $\pi$ enforces perfect destructive interference at $C_{n+1}$, thereby confining the particle within its AB cage.
Once the periodic drive is applied, new propagation channels appear via second-neighbor hoppings, as represented in Fig.~\ref{fig:secondneig}b and Fig.~\ref{fig:secondneig}c.
Crucially, because the driving imprints the same $\pi$-flux pattern on the secondary lattice, the destructive interference is not completely destroyed but rather \textit{redefined}.
In the example above, each pair of adjacent $C$ sites still exhibits perfect cancellation among all direct connecting paths (Fig.~\ref{fig:secondneig}b); however, a particle can now propagate coherently to the next-nearest $C_{n+2}$ site through combined first- and second-neighbor processes (Fig.~\ref{fig:secondneig}c).
In other words, the driving has established a new form of long-range connectivity within the lattice.

\begin{figure}
    \centering    \includegraphics{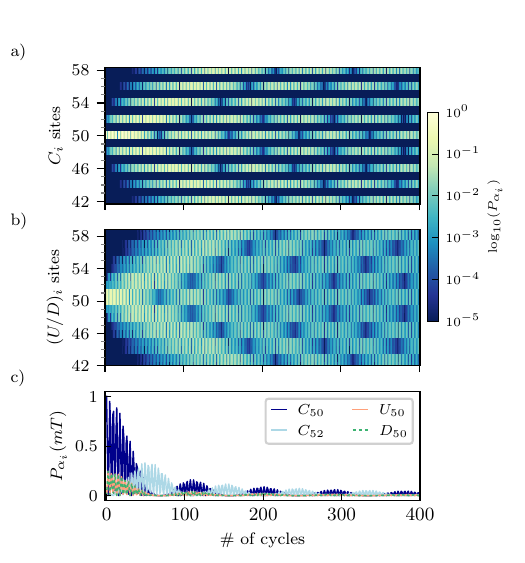}
    \includegraphics{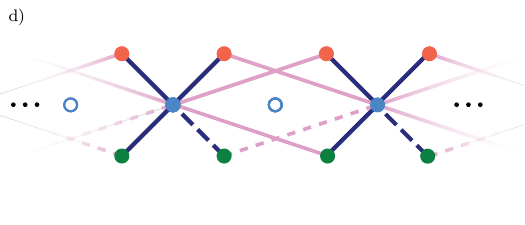}
    \caption{
    Stroboscopic evolution of the particle density $P_{\alpha_i}(mT) = \vert \langle  \alpha_i\vert \hat{U}(mT)\vert \psi_0\rangle \vert^2$ per lattice site as a
function of the drive cycles in \textbf{a)} central lattice sites, \textbf{b)} U (or equivalently D) lattice sites, and \textbf{c)} specific (labeled) lattice sites, as a
function of the drive cycles. Panels a and b are plotted in logarithmic scale. The initial state is $\ket{\psi_0}=\ket{C_{50}}$. The lattice onsite potentials are driven antisymmetrically with $\tilde{A} = 1.2$ and $\omega = 8$, $J = 1$, $N = 100$. \textbf{d)} Schematic representation of the effective lattice occupied by the particle, including the hoppings connecting the visible sites.}
\label{fig:dynamics_usublattice}
\end{figure}

To illustrate these ideas, we now analyze the stroboscopic real-space dynamics of a particle initially occupying a $C_i$ site of a finite-size closed chain, $\ket{\psi_0} = \ket{C_{50}}$. The site is chosen such that it is sufficiently far from the boundaries, to avoid the contribution from the edges, and we also consider a large enough length of the chain, with $N = 100$ unit cells (which translates into $3N + 1$ sites for the closed lattice). The results are shown in Fig. \ref{fig:dynamics_usublattice}, in which we plot the occupation probability $P_{\alpha_i}(mt) = \vert \langle  \alpha_i\vert \hat{U}(mT)\vert \psi_0\rangle \vert^2$ at stroboscopic times $mT$, where $m$ indicates the number of cycles. As it becomes evident in Fig.~\ref{fig:dynamics_usublattice}a, there is a clear destructive interference in alternating sites, while the $U/D$ lattices populate equally (see panel c in Fig.~\ref{fig:dynamics_usublattice}). As a function of time, the occupation of all sites exhibits a beating pattern (see panel c in Fig. \ref{fig:dynamics_usublattice}), due to the presence of nearly resonant frequencies contributing to the dynamics. This nearby frequencies appear due to the slight broadening of the original flatbands, which create finite detunings in the inter band transition energies. 

This dynamics shows the presence of an effective hopping network on the lattice in which one of two $C$ sites are removed, which means that the antisymmetric driving creates a \textit{dynamical selection rule}: the entire lattice effectively splits into two disconnected subsystems, as schematically shown in Fig. \ref{fig:dynamics_usublattice}, being all those unpopulated sites from the $C$ sublattice. This results from a striking interplay between driving and lattice symmetry, transforming AB caging from a purely local confinement mechanism into a tunable interference-assisted transport channel.

Let us note that, for a particle starting on a $U_n$ ($D_n$) site, the destructive interference condition migrates to the opposite $D$ ($U$) sublattice, selectively depleting alternating sites in one while populating the other (see Appendix \ref{sec:extra_dyn} for details on the dynamics starting from a different sublattice).\\

\paragraph*{\textbf{Decorated SSH chain.}} The dynamical selection rules in the driven chain can be analytically understood by considering the eigenvalues of the total effective Hamiltonian, up to second order: $\hat{H}_{\mathrm{eff}}(k) = \hat{H}_{\mathrm{eff}}^{(0)}(k) + \hat{H}_{\mathrm{eff}}^{(2)}(k)$. Since we now focus on the antisymmetric driving scheme, we drop the $\zeta = -1$ dependence in the Hamiltonian to simplify the notation. The overall structure of the effective Hamiltonian is 
\begin{equation}
    \hat{H}_{\mathrm{eff}}(k) = \left( 
    \begin{array}{ccc}
    0 & (*)& (*) \\
    J_1(1+e^{ik})+J_2(e^{-ik} + e^{2ik}) & 0 & 0 \\
    J_1^*(-1+e^{ik}) +J_2^*(e^{-ik} - e^{2ik})& 0 & 0
    \end{array} \right)
    \label{eq:h_eff_total} \ ,
\end{equation}
where $J_1$ is the total, effective first-neighbor hopping amplitude, 
\begin{equation}
    J_1 = \frac{i J }{\tilde{A} \pi}\left[ 1 - e^{i \tilde{A} \pi}\right] + \frac{10 \Gamma}{3\pi^3}\, \frac{J^3 \tilde{A} }{\omega^2}\sin\left( \pi \tilde{A} \right)\left[1 + e^{i \tilde{A} \pi}\right]
    \label{eq:j1}
\end{equation}
and $J_2$ is the second-neighbor hopping amplitude, 
\begin{equation}
    J_2 = \frac{2\Gamma}{\pi^3} \,\frac{J^3 \tilde{A} \, }{\omega^2}\sin\left( \pi \tilde{A} \right)\left[1 + e^{i \tilde{A} \pi}\right].
    \label{eq:j2}
\end{equation}
We refer the reader to Appendix \ref{sec:hopping_ren} for further details on the hopping amplitude renormalization. This Hamiltonian has exact eigenvalues
\begin{equation}
    E_{\mathrm{eff},\pm}(k) = \pm 2 \sqrt{\vert J_1 \vert^2 + \vert J_2 \vert^2 + 2 \vert J_1\vert \cdot \vert J_2 \vert \cos(2k)},
    \label{eq:bulk}
\end{equation}
and a flat band at $E_0 = 0$.  Interestingly, the band dispersion turns out to be equivalent to a SSH chain, with first- and second-neighbor hoppings representing its characteristic alternating pattern of hoppings (see Appendix \ref{app:decorated_SSH} for further details) and an enlarged unit cell. The geometry of the lattice provides a decoration to this effective SSH chain, resulting in the presence of the flatband at the center of the gap and two pairs of edge states, one in each gap. Previous works have analyzed the consequences of first-neighbor hopping dimerization in diamond lattices \cite{Bercioux2017, Martinez2024}. Our approach shows that this dimerization can be engineered indirectly through photo-induced long-range hopping processes.\\


Additionally, the numerical calculation of the Zak phase $\mathcal{Z}_{j} = i \int_{\text{FBZ}}dk\, \langle u_{j}(k)\vert \partial_t \vert u_{j}(k)\rangle$, where $\ket{u_{j}(k)}$ are the Floquet modes associated to each of the bands at stroboscopic times $mT$ ($m = 0,1,2,...)$, reveals that driving the system using our protocol does not modify the symmetries that keep $\mathcal{Z}_j$ quantized to their original values, due to the presence of the symmetries $\chi(k) = \mathrm{diag}\{1,-e^{ik},e^{ik}\}$ and $\hat{\Pi}(k) = \mathrm{diag}\{1,e^{ik},-e^{ik}\}$ \cite{Kremer2020} (see Appendix \ref{sec:symmetries}). \\

\paragraph*{\textbf{Dynamics in short arrays: the role of edge states.}}

\begin{figure}
    \centering
    \includegraphics{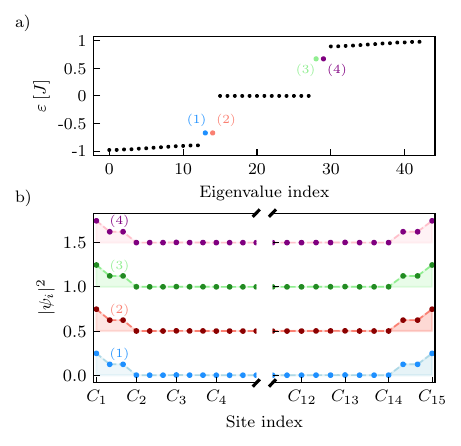}
    \caption{\textbf{a)} Quasienergies of the driven finite chain under asymmetric driving, with $N = 14$, $J = 1$, $\omega = 8$, $\tilde{A} = 1.2$, as in Figs. \ref{fig:quasienergies} and \ref{fig:dynamics_usublattice}. The edge states are highlighted in blue, red, green, and purple. \textbf{b)} Spatial profile of the edge state wavefunctions, as a function of the lattice site. They have shifted vertically for clearness. Their corresponding origin is indicated through the lower border of the shaded area.}
    \label{fig:driven_states}
\end{figure}

The case of a short array deserves separate consideration, as finite-size effects become dominant and the bulk region is strongly reduced. 
The most significant difference with respect to the bulk analysis lies in the role of the edge states in the dynamics. In the undriven finite lattice with an additional $C$-site termination, a particle initialized at the edge site $C_1$ cannot oscillate between boundaries via edge states, as typically occurs in conventional topological chains~\cite{BelloSublatticeDynamics, PhysRevLett.123.126401}. In the fully flat-band diamond chain, the edge states themselves are compact localized states, so a particle starting at $C_1$ remains permanently confined within the set ${C_1, U_1, D_1}$. The periodic driving changes this picture dramatically. By inducing long-range hopping processes, the drive opens new propagation channels that allow the particle to escape from its Aharonov–Bohm cage. For the edge states, this translates into the driving creating a finite overlap between extremeties of the lattice, but with a negligible leakage to the bulk. This is shown in Fig. \ref{fig:driven_states}.

\begin{figure}
    \centering
    \includegraphics[width = \linewidth]{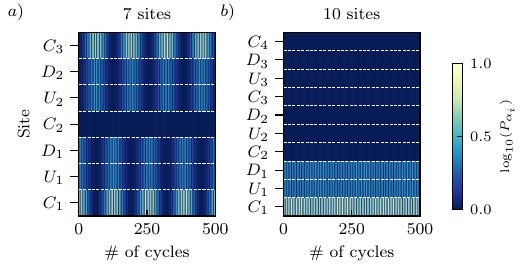}
    \caption{
    Stroboscopic evolution of the particle density $P_{\alpha_i}(mT) = \vert \langle  \alpha_i\vert \hat{U}(mT)\vert \psi_0\rangle \vert^2$ per lattice site as a
function of the drive cycles for short lattices with \textbf{a)} even number of cells $N = 2$ (7 sites)  and \textbf{b)} odd number of cells $N = 3$ (10 sites). The initial state is $\ket{\psi_0}=\ket{C_{1}}$. The lattice onsite potentials are driven anti-symmetrically with $\omega = 8$, $\tilde{A} = 1.2$, $J = 1$. The white, dashed lines serve as guidance to identify correctly the occupation of each site.}
    \label{fig:dynamics_shortchain}
\end{figure}

Interestingly, the system size also plays a crucial role: one out of every two $C$ sites becomes effectively depleted in the particle dynamics due to the interference conditions discussed above. As a result, for even $N$, the particle can reach the opposite end of the chain, while for odd $N$, destructive interference prevents this transfer. This behavior is illustrated in Fig.~\ref{fig:dynamics_shortchain}, where we analyze the dynamics of a particle in an array of $N=2$ and $N=3$ unit cells (corresponding to $3N+1 = 7$ and $10$ sites, respectively). The initial state is $\ket{\psi_0}=\ket{C_1}$, and again, we plot the occupation probability of each site $P_{\alpha_i}(mT)$ for stroboscopic times. For $N=2$, the particle reaches the opposite edge and undergoes coherent oscillations between the two terminations, whereas for $N=3$, it remains trapped within its initial plaquette. Let us note that a small leakage is to be expected in $C_2$ for $N = 2$, as well as a negligible occupancy of $C_4$ for $N = 3$ at larger stroboscopic times, both of them consequences of finite-size effects.

\section{Entanglement protocol in a small lattice \label{sec:entanglement}}

\paragraph*{\textbf{Entanglement measurements.}} As demonstrated above for a single-particle Hamiltonian, a sufficiently high-frequency driving  of the ABF diamond lattice preserves compact localization, as shown by $\hat{H}^{(0)}_{\mathrm{eff}}(k)$ in Eq. \eqref{eq:h_eff0}. Reducing the frequency activates long-range hopping,  breaking the compact localization of the states and yielding an effective Hamiltonian of the form of Eq. \eqref{eq:h_eff_total}. Now, building on this analysis, we outline a protocol that harnesses the antisymmetric drive to generate two-particle entanglement. 

As a minimal setting, we consider a 7-sites system composed of two complete unit cells and an additional $C$ site, forming a double plaquette as in Fig. \ref{fig:dynamics_shortchain}a. As shown above, such a geometry ensures particle transfer across the chain, mediated by the edge states. As compared to the single-particle case discussed above, we also add a first-neighbor density-density interaction term, 
\begin{eqnarray}
    \hat{H}_{\mathrm{int}} & = & V\sum_n^{N}  \left( \hat{N}_{U,n } \hat{N}_{C,n } + \hat{N}_{D,n} \hat{N}_{C,n } \nonumber \right.\\
    & & \quad \quad \quad \quad  \left. + \hat{N}_{U,n } \hat{N}_{C,n + 1 } + \hat{N}_{D,n } \hat{N}_{C,n+1 } \right) \ ,
\end{eqnarray}
where $\hat{N}_{\alpha,n} = \sum_{\sigma = \uparrow,\downarrow}\hat{\alpha}^\dagger_{n,\sigma} \hat{\alpha}_{n,\sigma}$,with $\alpha = C, U, D$. 

Let us consider two same-spin fermions occupying opposite ends of the double plaquette at $t = 0$, i.e., $\ket{\Psi_0} = \hat{C}^\dagger_{1} \hat{C}^\dagger_{3} \ket{0}$ (the spin component has been dropped for simplicity). In the high-frequency regime, long-range hoppings are suppressed and the particles will remain unentangled, as bounded as they are to their AB cages, regardless of the presence of the drive (in fact, for the purpose of the protocol, the high-frequency regime is no different from the undriven chain). By tuning the time-periodic field to a lower frequency, however, quasi-flat bands with small gaps activate, and so does transport across the lattice. As we will now show, the coherent interference in the driven system can build up entanglement  if the particles interact. At a stroboscopic time with maximal entanglement between spatially separated particles, the drive is switched back to the high-frequency regime, which suppresses further tunneling, restores compact localization, and freezes the entangled state for readout.

To quantify spatial entanglement along the chain, we introduce a bipartition in which subsystem
$A = \{ C_1, U_1, D_1, C_2 \}$ contains the left end of the structure and $B = \{ U_2, D_2, C_3\}$ contains the right end. A natural diagnostic is the von Neumann entropy $S^{(A)} = - \mathrm{Tr}\{ \rho^{(A)} \log (\rho^{(A)})\}$, which measures how strongly the wavefunction spreads across the cut. However, $S^{(A)}$ does not directly characterize two-party entanglement: particle delocalization or temporary exchange across the boundary of the cut can increase $S^{(A)}$ even when the system ultimately returns to a configuration with one particle per region and the final state factorizes into a product of two localized fermions.

To isolate the accessible or operational entanglement, i.e., the part that remains meaningful under local particle-number superselection, we restrict to the sector with one particle in each subsystem, $n_A = n_B = 1$. We first project the density matrix onto the corresponding occupation sector, 
\begin{equation}
    \hat{\rho}_{1,1} = \frac{\hat{P}_{1,1} \, \hat{\rho} \, \hat{P}_{1,1}}{p_{1,1}},\quad p_{1,1} =  \mathrm{Tr}\{ \hat{P}_{1,1} \, \hat{\rho} \, \hat{P}_{1,1}\} \ ,
\end{equation}
where $P_{\alpha,\beta}$ projects onto the subspace $n_A = \alpha$ and $n_B = \beta$ subspace. The Von Neumann entropy of the reduced density matrix $\hat{\rho}^{(A)}_{1,1}$ then gives $S^{(A)}_{1,1} = - \mathrm{Tr}\{ \rho^{(A)}_{1,1} \log \rho^{(A)}_{1,1} \}$. The operational entanglement is defined as 
\begin{equation}
    E_P(1,1) = p_{1,1} \, S^{(A)}_{1,1} \ .
    \label{eq:op_entanglement}
\end{equation}

In summary, $S^{(A)}$ captures all quantum correlations across the bipartition, while $E_P(1,1)$ filters out contributions arising from mere particle delocalization. If $S^{(A)}$
grows but $E_P(1,1)$ vanishes, the correlations are transient and nonoperational. A finite 
$E_P(1,1)$ once the particles are well separated signals that the process has generated genuine, shareable two-particle entanglement between the two particles.\\

\begin{figure}
    \centering
    \includegraphics{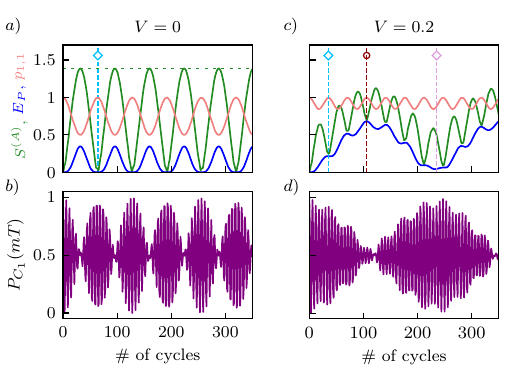}
    \caption{\label{fig:entanglement} Stroboscopic evolution of the entropy of the reduced density matrix $S^{(A)}(mT)$, the operational entanglement $E_P(1,1)(mT)$, the probability of one particle per sector $p_{1,1}(mT)$, for both \textbf{a)} $V = 0$, and \textbf{c)} $V = 0.2$. The dashed, horizontal line in panel a) shows the maximum value for $S^{(A)}$. The vertical, dashed blue and purple lines represent the oscillation periods of $S^{(A)}$ and $E_P(1,1)$. The vertical, dashed line in red corresponds to the special point at which $S^{(A)} = E_P(1,1)$. Panels \textbf{b)} and \textbf{d)} show the particle density in $C_1$ site as a function of the drive cycles for a 7-site lattice initialized in state $\ket{\Psi_0}=\hat{C}_1^{\dagger} \hat{C}_3^{\dagger}\ket{0}$, for $V = 0$ and $V = 0.2$, respectively. The lattice onsite potentials are driven antisymmetrically with $\tilde{A} = 1.2$ and $\omega = 8$ ($J = 1$).}
\end{figure}

\paragraph*{\textbf{Numerical results.}} To build intuition, we first analyze the noninteracting limit $V=0$. In this case the two-particle state remains a Slater determinant, so no genuine fermionic correlations are expected. Figure~\ref{fig:entanglement} shows the stroboscopic evolution of $S^{(A)}(mT)$ and $E_P(1,1)(mT)$, both displaying pronounced oscillations. These arise from the periodic buildup of spatial entanglement as the particles spread along the chain: starting from a fully separated configuration ($S^{(A)} = 0$), the wavefunctions delocalize across the partition $A|B$, reaching the maximal bipartite entropy $S^{(A)} = 2\ln2$ (horizontal, dashed line in the plot), which is consistent with the character of the boundary modes engineered by the drive~\cite{PhysRevB.73.245115, PhysRevA.101.052323}. When the particles later return to opposite sides, the spatial entanglement collapses again to zero, $S^{(A)} = 0$. Only part of this spatial entanglement is operationally accessible, as quantified by $E_P(1,1)(mT)$. The curve $p_{1,1}(mT)$ also deserves attention on its own. The behavior of the sector weight $p_{1,1}(mT)$ is particularly revealing: the condition $p_{1,1}\equiv 1$  identifies times when the particle number across the cut is sharp, implying, implying $E_P(1,1) = S^{(A)}$. In the present noninteracting case, this occurs only at the minima, where $S^{(A)} = E_P(1,1) = 0$. Thus, whenever the particles fully separate into the two sectors, all entanglement vanishes. In the absence of spatial overlap across the bipartition, no operational entanglement can be generated.

To further characterize the two-particle dynamics, we also show the occupation probability of the leftmost site, $P_{C_1}(mT)$, evaluated at stroboscopic times. This quantity reveals that nearly all of the initial population at the chain ends returns to them, with only a small leakage into the bulk due to finite-size effects. As anticipated, the dynamics is governed by the drive-induced edge states (see Fig. \ref{fig:dynamics_shortchain}). The oscillation frequency of $S^{(A)}(mT)$ and $E_P(1,1)(mT)$ coincides with the frequency of maximums in $P_{C_1}(mT)$ (vertical dashed blue line). Notably, this frequency is only an effective one: the beating pattern in $P_{C_1}$indicates the presence of two nearby frequencies, which originate from the small energy splittings within each pair of edge states in the system.
When interactions are switched on, the same protocol generates genuine fermionic correlations. Over longer timescales the system develops a significantly larger operational entanglement than in the noninteracting case. As compared to the $V = 0$ case, there are two characteristic frequencies in the oscillations of both $S^{(A)}(mT)$ and $E_P(1,1)(mT)$. This shift originates from interaction-induced mixing of nearby quasienergy levels, which reduces the effective band gaps. As a result, interactions favor slower oscillations that dominate the long-time entanglement dynamics (see Appendix \ref{sec:FT_entanglement} for more details). Crucially, we identify a stroboscopic time at which $p_{1,1} = 1$ while $S^{(A)} = E_P(1,1) \neq 0$, indicating that the two particles form an entangled state while being spatially separated across the bipartition (dashed–dotted red line). As above, this point corresponds to a $P_{C_1} \approx P_{C_3}  \approx 0.5$ (only $P_{C_1}$ shown in the plot). Returning to the high-frequency regime at this moment suppresses further tunneling, effectively freezing the entangled state for readout. 

These results demonstrate that the interplay between lattice geometry and drive-induced hopping patterns enables the controlled generation of two-particle entanglement. The protocol thus allows both the creation and the storage of spatially separated entangled states.

\section{Conclusions \label{sec:conclusions}}

We have shown that Floquet engineering provides a powerful route to preserve, enhance, and fundamentally reshape flat-band phenomena. By periodically driving the on-site energies of the 
$U/D$ sites of an ABF diamond lattice, we demonstrated that the connectivity of the lattice can be dynamically reconfigured: a static system hosting compact localized states is converted into one supporting tunable quasi-flat bands, while a residual flat mode remains pinned at zero energy. In the high-frequency regime, the drive generates an emergent \textit{secondary} diamond lattice, spanning two original plaquettes, that preserves the $\phi = \pi$ condition. This effective connectivity arises from the interplay between geometry and drive-induced symmetries, producing new tunneling pathways that lift the static Aharonov–Bohm caging conditions and enable controlled delocalization and long-range correlations. As a result, robust edge modes appear with minimal leakage into the bulk. Unlike in the static lattice, the induced dispersive character cannot be reversed by adjusting $\phi$ alone, reflecting its intrinsic link to the presence of the driving field.\\

Beyond their fundamental interest, these driven quasi-flat bands offer a versatile mechanism for generating and manipulating quantum correlations in lattice systems. The driven quasi-flat bands enable coherent charge oscillations across distant sites using the boundary modes \cite{PhysRevResearch.2.033475, PhysRevA.93.032310}, allowing for the controlled generation of two-particle entanglement that could be probed in current experimental platforms \cite{Chen2025, Kiczynski2022}. More broadly, our results indicate that periodically modulated flat-band lattices constitute dynamically reconfigurable architectures for quantum information processing, where geometry, interference, and driving symmetries can be jointly exploited to encode and transport quantum correlations. The non-zero, quantized invariants $\mathcal{Z}_j$ identified here further suggest a degree of topological robustness in the transport process, providing an added layer of reliability and tolerance against imperfections \cite{Lang2017, PhysRevResearch.2.033475}.

\section*{Acknowledgements}
MB and AA acknowledge funding from the Emmy Noether Programme of the German Research Foundation (DFG) under grant no. BE 7683/1-1. 

\appendix

\section{Edge states in the closed vs open static chain \label{app:edgestates}}

\begin{figure}
    \centering
\includegraphics{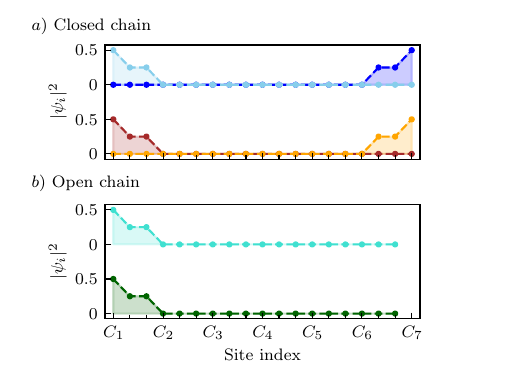}
    \caption{ Edge state wavefunction $\vert \psi_i\vert^2$ as a function of the lattice site, for \textbf{a)} the closed chain depicted in Fig. \ref{fig:schematic} with $N = 6$ unit cells, $J = 1$, and $\phi = \pi$, \textbf{b)} the open chain with the same $N$. The sites are ordered as $\{ C_1, U_1, D_1,C_2, U_2, D_2,...\}$.}
    \label{fig:edge_static}
\end{figure}

In this section we analyze the edge states appearing in the static, diamond chain, for both open and closed boundary. Fig. \ref{fig:edge_static} shows the spatial profile $\vert \psi_i \vert^2$ of the edge states in a closed chain, panel a, versus open chain, panel b, for a finite-size system with $N = 6$ unit cells. The edge states in the second gap are shifted vertically for clearness in both cases. In the closed case, each gap hosts two edge states, localized at each of the ending $C$ sites of the chain. In the open case, there is one edge state per gap, with support in only one end of the chain. For all of them, the edge states are compact localized states, with no bulk decay. 

The energy for such states can be obtained by seeking an eigenstate $\psi$ of the Hamiltonian with support only on the first cell $(C_1, U_1, D_1)$, and decoupled from $C_2$: $\psi = (\psi_{C_1}, \psi_{U_1}, \psi_{D_1}, 0,0,...,0)^T$. Because $C_2$ sees $U_1$ and $D_1$ with the same sign $J$, the decoupling condition is $\psi_{U_1} + \psi_{D_1} = 0$. Writing the eigenvalue equations on the first cell gives 
\begin{eqnarray}
    E \, \psi_{U_1} & = & J \, \psi_{C_1},\\
    E \, \psi_{D_1} & = & - J \psi_{C_1},\\
    E \, \psi_{C_1} & = & J \psi_{U_1} - J \psi_{D_1} = 2J \psi_{U_1}.
\end{eqnarray}
Eliminating $\psi_{C_1}$ using the first line, $\psi_{C_1} = (E / J) \psi_{U_1}$, and plugging in back into the third, we get the energy of the edge state,
\begin{equation}
    E_{\mathrm{e.s.}} = \pm \sqrt{2} J.
\end{equation}
A normalized left-edge eigenvector is 
\begin{equation}
    \ket{ \psi^L_{\pm}} = \frac{1}{2}\left( \ket{U_1} - \ket{D_1} \pm \sqrt{2} \ket{C_1} \right),
\end{equation}
which is compact (strictly zero on all other sites). By symmetry, the right edge has an analogous compact state on the last cell; with the same termination it also appears at the same energy. 

For the purpose of this work, it is also enlightening to look into the bulk eigenvectors corresponding to the zero-energy band, $E_0 = 0$. From $\hat{H}_{\mathrm{rhomb}}(k)\psi = 0$, we obtain that the only constraint is $\psi_{C_1} = 0$, that is, the states do not couple to any $C$ site. We can choose a normalized combination of $\ket{U_k}$ and $\ket{D_k}$ orbitals, such that
\begin{equation}
    \ket{\psi_0(k)} = \frac{1}{\sqrt{2}}\left( \ket{U_k} + \ket{D_k} \right).
\end{equation}
In real space, this corresponds to a compact localized state spanning only sites belonging to the majority sublattice $\{ U_n, D_n\}$.

\section{Static on-site energies: effect on the band structure \label{app:onsite_energies_static}}

Let us first analyze the case of identical onsite energy in both $U$ and $D$ sites, $\epsilon_u = \epsilon_d = \epsilon$, 
\begin{equation}
    \hat{H}_{0}(k) = J \left( \begin{array}{ccc}
    0 & t_u^*(k) & t_d^*(k)\\
    t_u(k) & \epsilon & 0\\
    t_d(k) & 0 & \epsilon
    \end{array} \right)
    \label{eq:general_hamiltonian}
\end{equation}
where we have defined $t_u(k) = 1 + e^{ik}$ and $t_d(k) = e^{i\phi} + e^{ik}$ for simplicity. We now build the symmetric and antisymmetric combinations on the $U/D$ sublattice, 
\begin{eqnarray}
    \ket{S_k} &  = & \frac{1}{\vert t(k) \vert}\left( t_u(k)\ket{U_k} + t_d(k)\ket{D_k} \right), \label{eq:symmetric_comb}\\
    \ket{A_k} &  = & \frac{1}{\vert t (k) \vert}\left( - t_d^*(k)\ket{U_k} + t_u^*(k)\ket{D_k} \right), \label{eq:asymmetric_comb}
\end{eqnarray}
with $\vert t (k) \vert = \sqrt{\vert t_u(k)\vert^2 + \vert t_d(k)\vert^2} = \sqrt{2} \sqrt{2 + \cos(k)+\cos(k-\phi)}$, and transform $\hat{\mathcal{H}}_{\mathrm{static}}(k)$ to the basis of $\{ \ket{C_k}, \ket{S_k}, \ket{A_k} \}$, to get
\begin{equation}
    \hat{H}^\prime_0(k) = \left( \begin{array}{ccc}
    0 & \vert t(k) \vert & 0 \\
    \vert t(k) \vert & \epsilon & 0 \\
    0 & 0 & \epsilon
    \end{array}
    \right).
    \label{eq:h_after_U1}
\end{equation}
In this new basis, a $C$ orbital only couples to a $S$ orbital, with strength $\vert t(k) \vert$, while being decoupled from a $A$ site. The antisymmetric combination yields precisely an eigenstates for every $k$: a flatband at energy $E_0 = \epsilon$. Therefore, the central flatband is preserved, though shifted by $\epsilon$. The diagonalization of the first block yields the dispersive pair, 
\begin{equation}
    E_{\pm}(k) = \frac{1}{2}\left( \epsilon \pm \sqrt{\epsilon^2 + 4\vert t(k)\vert^2}\right).
\end{equation}
For $\phi = \pi$, in fact, we find that all bands remain flat, with $E_{\pm} = \frac{1}{2}\left( \epsilon \pm \sqrt{\epsilon^2 + 16J^2}\right)$.\\

In the antisymmetric case, the transformation yields a different result. In fact, the transformation to the basis $\{ \ket{C_k}, \ket{S_k}, \ket{A_k} \}$ does not decouple the orbitals, 

\begin{equation}
	\hat{H}_0^\prime(k) = \left(
	\begin{array}{ccc}
		0 & \vert t(k) \vert & 0 \\
		\vert t(k) \vert & \epsilon \left( 1 - \frac{2 \vert t_d(k) \vert^2}{\vert t(k) \vert^2} \right) & \frac{2\epsilon t_d^*(k) t_u^*(k)}{\vert t(k) \vert^2} \\
		0  &  \frac{2\epsilon t_d(k) t_u(k)}{\vert t(k) \vert^2} &  - \epsilon \left( 1 - \frac{2 \vert t_d(k) \vert^2}{\vert t(k) \vert^2} \right)
	\end{array}
	\right).
\end{equation}

Only for $\phi = 0$, $t_u(k) = t_d(k)$ and all diagonal entries of $\hat{H}_0^\prime(k)$ are cancelled out, and the system is allowed to host a flatband at $E_0 = 0$. For $\phi \neq 0$, all bands are dispersive. 

\section{Effective Hamiltonian}

\subsection{High-frequency expansion \label{app:effective_hamiltonian}}

After rotating the Hamiltonian into the frame of the driving field and obtaining $\hat{H}^\prime_{\mathrm{tot}}(k,t)$ in Eq. \eqref{eq:transformed_H}, we apply van Vleck perturbation theory to obtain $\hat{H}_\text{eff}$ up to second order in $1/\omega$,
\begin{subequations}
    \begin{align}
    \hat{H}^{(0)}_\text{eff}(k,t) & = \hat{H}_0 =  \frac{1}{T} \int_0^T \hat{H}^\prime_{\mathrm{tot}}(k,t)\, dt \\
    \hat{H}^{(1)}_\text{eff}(k,t) &= \frac{1}{\omega} \sum_{l=1}^\infty \frac{1}{2l} \big[\hat{H}_l, \hat{H}_{-l}\big], \\
    \nonumber \hat{H}^{(2)}_\text{eff}(k,t) &= \frac{1}{\omega^2} \sum_{m\neq 0} \left( \frac{1}{2m^2} \big[\big[\hat{H}_{-m}, \hat{H}_0\big], \hat{H}_m\big] \right.\\
    &\left. \quad \quad \quad + \sum_{l'\neq 0, l} \frac{1}{3ll'} \big[\big[\hat{H}_{-l}, \hat{H}_{l-l'}\big], \hat{H}_{l'}\big] \right). \label{eq:h2}
\end{align}
\end{subequations}
where the $\hat{H}_n$ are the harmonics of the Fourier expansion, 
\begin{equation}
    \hat{H}_n = \frac{1}{T}\int dt \, \hat{H}^\prime_{\mathrm{tot}}(k,t) e^{in\omega t}.
\end{equation}
While the zeroth-order term is already shown in Eqs. \eqref{eq:effective_h} and \eqref{eq:effective_J0}, the first order cancels out. Therefore, higher order corrections come necessarily from Eq. \eqref{eq:h2}, for both symmetric and asymmetric driving.\\

\subsection{Effective Hamiltonian for symmetric driving \label{sec:symmetric}}

While the resulting $\hat{H}_{\mathrm{eff}}(k,t)$ for $\zeta = -1$ is discussed in the main text, here we also include the explicit result for $\hat{H}^{(S)}_{\mathrm{eff}}(k,t)$ for symmetric driving, $\zeta = +1$. The zeroth-order term only differs in the sign of the hoppings $C \leftrightarrow D$, as compared to Eq. \eqref{eq:h_eff0}, 
\begin{equation}
    \hat{H}_{\mathrm{eff}}^{(0,S)}(k) = \left( 
    \begin{array}{ccc}
    0 & (*)& (*) \\
    J_0(1+e^{ik}) & 0 & 0 \\
    J_0(-1+e^{ik})& 0 & 0
    \end{array} \right).
\end{equation}
The second-order term has the following structure, 
\begin{equation}
    \hat{H}_{\mathrm{eff}}^{(2,S)}(\zeta = +1) = \left( \begin{array}{ccc}
        0 & (*) & (*) \\
        \mathcal{C}^{S} \, h^{(S)}_{cu}(k) & 0 & 0 \\
        \mathcal{C}^{S} \, h^{(S)}_{cu}(k) & 0 & 0
    \end{array}\right)
\end{equation}
where the coefficient $\mathcal{C}^S$ is given by
\begin{eqnarray}
    \mathcal{C}^{S} & = & \frac{16 \Gamma}{3\pi^3} \,\frac{J^3 \tilde{A}}{\omega^2} \, \sin\left(\pi \tilde{A} \right)\left[1 + e^{ i \tilde{A} \pi }\right]
\end{eqnarray}
where $\Gamma$ has the same definition as in Eq. \eqref{eq:gamma_Def}, and the dispersive part is shown in Eq. \eqref{eq:dispersive_part}.\\

\subsection{Convergence properties of $\Gamma$ \label{sec:convergence}}

\begin{figure}
    \centering
    \includegraphics{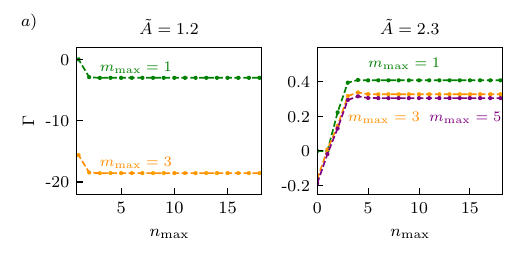}
    \includegraphics{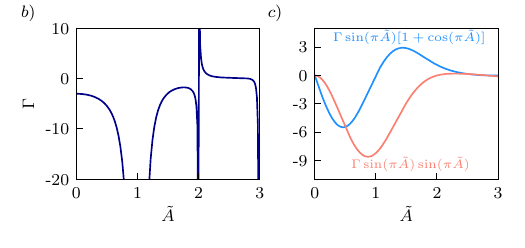}
    \caption{Convergence properties of $\Gamma$. \textbf{a)} Value of $\Gamma$ as a function of the maximum cutoff $n_{\mathrm{max}}$, for different values of $m_{\mathrm{max}}$, until convergence is obtained, for both $\tilde{A} = 1.2$ (used in all plots in the main text), and $\tilde{A} = 2.3$. \textbf{b)} Value of $\Gamma$ after convergence, as a function of $\tilde{A}$. $\Gamma$ diverges at integer values of $\tilde{A}$, but together with the additional dependence on $\tilde{A}$ shown in Eqs. \eqref{eq:j1} and \eqref{eq:j2}, the total hopping renormalization is well defined for arbitrary values of $\tilde{A}$. }
    \label{fig:convergence}
\end{figure}

Now, we analyze the convergence properties of $\Gamma$. Fig. \ref{fig:convergence}a shows the value of $\Gamma$ for different choices of the maximum cutoff $n_{\mathrm{max}}$, $m_{\mathrm{max}}$, and how convergence is easily obtained after adding the first harmonics to the series expansion. In panel \ref{fig:convergence}b we plot the behavior of $\Gamma$ as a function of $\tilde{A}$. Due to its pole structure with $\Pi_\alpha = (\tilde{A}^2 - \alpha^2)$ ($\alpha = n,m,n-m$), Eq. \eqref{eq:gamma_Def}, it diverges at integer $\tilde{A}$, however the total photo-dressed hopping amplitudes do not, Eqs. \eqref{eq:j1} and \eqref{eq:j2}, as they also contain the additional factors $\sin(\pi \tilde{A})[1 + \mathrm{exp} ( i\tilde{A} \pi )]$. Panel \ref{fig:convergence}c shows this: both the real (blue) and imaginary (red) part of $\Gamma \sin(\pi \tilde{A})[1 + \mathrm{exp} ( i\tilde{A} \pi )]$ are well behaved at integer $\tilde{A}$.

\begin{figure}
    \centering
    \includegraphics{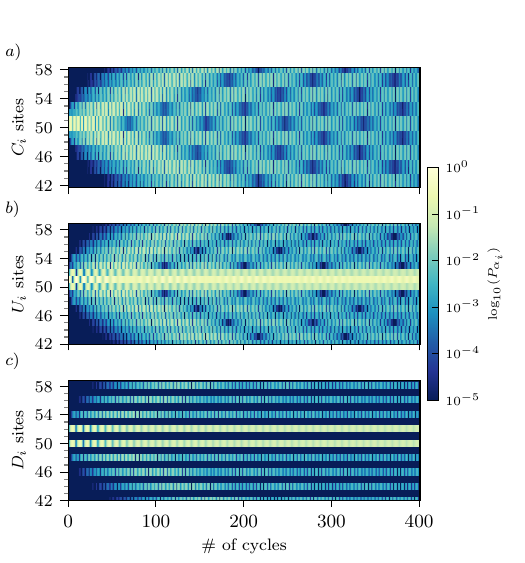}
    \caption{Stroboscopic evolution of the particle density $P_{\alpha_i}(kt) = \vert \langle  \alpha_i\vert \hat{U}(mT)\vert \psi_0\rangle \vert^2$ in \textbf{a)} central lattice sites, \textbf{b)} U lattice sites, and \textbf{c)} U lattice sites, as a
function of the drive cycles. The occupation is plotted in logarithmic scale. Notice that the occupation probability is plotted in logarithmic scale. The initial state is $\ket{\psi_0}=\ket{U_{50}}$. The lattice onsite potentials are driven antisymmetrically with $\tilde{A} = 1.2$, and $\omega = 8$ ($J=1$)}
    \label{fig:ulattice}
\end{figure}

\section{Additional results on charge dynamics \label{sec:extra_dyn}}

We now turn to the real-space dynamics of a particle initially localized at $\ket{\psi_0} = \ket{U_{50}}$, considering again a chain with $N = 100$ sites, as in Fig.~\ref{fig:dynamics_usublattice}. The corresponding results are presented in Fig.~\ref{fig:ulattice}, which shows the occupation probability $P_{\alpha_i}(mT) = |\langle \alpha_i | \hat{U}(mT) | \psi_0 \rangle|^2$ at stroboscopic times, plotted on a logarithmic scale. The destructive interference condition is now transferred to the $D$ sublattice, resulting in an alternating occupation pattern along the $D$ sites.

\section{Hopping renormalization in the effective model \label{sec:hopping_ren}}

\begin{figure}
    \centering
    \includegraphics{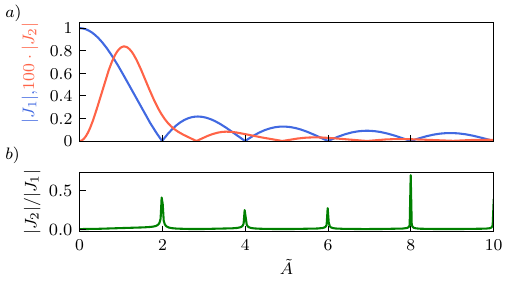}
    \caption{\textbf{a)} $|J_1|$ and $|J_2|$ as a function of $\tilde{A}$, with $J = 1$ and $\omega = 8$. \textbf{b)} Ratio between hopping amplitudes, for the same choice of parameters.}
    \label{fig:hopping_ren}
\end{figure}

Let us now examine in detail the hopping renormalization of the effective Hamiltonian under asymmetric driving. The drive introduces complex hopping phases that depend on the parameter $\tilde{A}$. These phases, however, are \textit{gauge-trivial} and can be removed by a local $U(1)$ transformation of the form $\hat{\alpha}_n \rightarrow \hat{\alpha}_n e^{-i\varphi_{\alpha_n}}$, with $\alpha = \{C,U,D\}$. Let us write Eqs. \eqref{eq:j1} and \eqref{eq:j2} as $J_1 = \vert J_1 \vert e^{i\varphi_1}$ and $J_2 = \vert J_2 \vert e^{i\varphi_2}$. A simple proof of how the gauge transformation works can be outlined by requiring, for the upper arm of the diamond chain and spanning two cells, the following conditions, 
\begin{eqnarray}
    \varphi_{U_1} - \varphi_{C_1} & = & -\varphi_1, \\
    \varphi_{C_2} - \varphi_{U_1} & = &  \varphi_1, \\
    \varphi_{U_2} - \varphi_{C_2} & = &  - \varphi_1, \\
    \varphi_{C_3} - \varphi_{U_2} & = &  \varphi_1, \\
    \varphi_{C_1} - \varphi_{U_2} & = &  \varphi_2,
\end{eqnarray}
where the phases are consistent with those derived in Eq.~\eqref{eq:h_eff_total}.
This set of equations can be solved in terms of $\varphi_1$, $\varphi_2$, and an arbitrary reference phase $\varphi_{C_1}$, for any pair $\{\varphi_1, \varphi_2\}$. Hence, the complex phases generated by the driving are removable through a local gauge choice, and the physical properties of the system are equivalent to those of a real Hamiltonian with hopping amplitudes $|J_1|$ and $|J_2|$. The value of $|J_1|$ and $|J_2|$ as a function of $\tilde{A}$ is shown in Fig. \ref{fig:hopping_ren}a.

The only phases that cannot be gauged away are those associated with the $\pi$ flux threading the original plaquettes, and the identical flux imprinted onto the emergent secondary diamond lattice formed by second-neighbor hoppings.
This residual $\pi$ phase manifests as the overall minus sign in the dispersive part of the spectrum, Eqs. \eqref{eq:hoppings_ctou} and \eqref{eq:hoppings_ctod}.

\section{Connection with the SSH Hamiltonian in $k$ space \label{app:decorated_SSH}}

In this section, we derive step by step the mapping of $\hat{H}_{\mathrm{eff}}(k)$ in Eq. \eqref{eq:h_eff_total} to a SSH chain. First of all, as shown in Appendix \ref{app:onsite_energies_static}, any Hamiltonian with the structure of Eq. \eqref{eq:general_hamiltonian} can be taken to block-diagonal form by using a symmetric/asymmetric combination of $U/D$ orbitals. Doing this transformation on $\hat{H}_{\mathrm{eff}}(k)$, with 
\begin{eqnarray}
    t_u(k) & = & J_1(1+e^{ik})+J_2(e^{-ik} + e^{2ik}),\\
    t_d(k) & = & J_1^*(-1+e^{ik}) +J_2^*(e^{-ik} - e^{2ik}),
\end{eqnarray}
gives a decoupled block of $E_0 = 0$ (flat band), and a $2\times 2$ block with off-diagonal terms
\begin{equation}
    \hat{H}_{\mathrm{block}}(k) = \left( \begin{array}{cc}
    0 & \vert t(k) \vert \\
    \vert t(k) \vert & 0
    \end{array} \right)
\end{equation}
where
\begin{equation}
    \vert t(k) \vert = 2\sqrt{\vert J_1 \vert ^2 + \vert J_2\vert^2 + 2 \vert J_1 \vert \cdot \vert J_2 \vert \cos(2k)},
    \label{eq:disp_ssh}
\end{equation}
as already shown in Eq. \eqref{eq:h_after_U1}. Notice that the phase of the hoppings is irrelevant, as everything is written in terms of the modulus of both $J_1$ and $J_2$. In the following, and for the shake of clearness, we drop the $\vert \cdot \vert $ and write simply $J_1$ and $J_2$.

Clearly, the eigenvalues of this block are given by $E_{\pm}(k) = \pm \vert t(k)\vert$. The connection to the SSH chain is already justified by direct comparison of Eq. \eqref{eq:disp_ssh} to the standard bulk bands of a dimerized chain,
\begin{equation}
    E_{\pm, \mathrm{SSH}} = \pm \sqrt{t_1^2 + t_2^2 + 2t_1 t_2 \cos(k)}
\end{equation}
with a reduced Brillouin zone and scaled hoppings, $t_1 \rightarrow 2J_1 $ and $t_2 \rightarrow 2 J_2$. However, we still can bring the $\hat{H}_{\mathrm{eff},2\times 2}$ into a SSH-like structure, by considering a rotation $\hat{R}(k)$ with the basis
\begin{eqnarray}
    \vert \phi_{\pm} \rangle & = & \frac{1}{\sqrt{2}}\left( 
    \begin{array}{c}
        \pm \frac{\sqrt{J_1^2 + J_2^2 + 2J_1 J_2 \cos(2k)}}{J_1 + e^{2ik}J_2}\\
        1
    \end{array} \right),
\end{eqnarray}
as $\hat{R}(k) = \left( \ket{\phi_-} \vert \ket{\phi_+} \right)$. In total, the transformed Hamiltonian yields
\begin{eqnarray}
    \hat{R}(k) \, && \hat{R}_z^{-1} \, \hat{H}_{\mathrm{block}}(k) \hat{R}_z \hat{R}(k)^{-1} \nonumber \\ &&= \left( \begin{array}{cc}
        0 & -2J_1 - 2J_2 e^{-2ik} \\
        -2J_1 - 2J_2 e^{-2ik} & 0
    \end{array} \right)
    \label{eq:h_sshform}
\end{eqnarray}
where $\hat{R}_z$ is nothing but the rotation from $x$ to $z$ to bring $\hat{H}_{\mathrm{block}}$ into diagonal form, 
\begin{equation}
    \hat{R}_z = \frac{1}{\sqrt{2}}\left( \begin{array}{cc}
    1 & 1 \\
    1 & -1
    \end{array} \right).
\end{equation}

\begin{figure}
    \centering
    \includegraphics{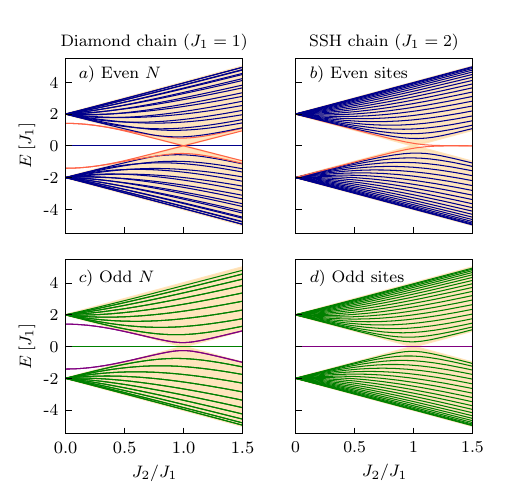}
    \caption{Comparison between the energy spectrum of: \textbf{a), c)} a diamond chain with first- and second-neighbor hoppings, with even $N = 20$ and odd $N = 21$; \textbf{b), d)} a SSH chain, with even ($2N$) and odd ($2N + 1$) number of sites, for $N = 20$.}
    \label{fig:diamond_vs_ssh_bulk}
\end{figure}

Figure~\ref{fig:diamond_vs_ssh_bulk} compares the energy spectrum of a diamond chain with first- and second-neighbor hoppings $J_1$ and $J_2$, as described by the effective Hamiltonian in Eq.~\eqref{eq:h_eff_total} (left column, panels \ref{fig:diamond_vs_ssh_bulk}a and \ref{fig:diamond_vs_ssh_bulk}c), with that of an SSH chain (right column, panels \ref{fig:diamond_vs_ssh_bulk}b and \ref{fig:diamond_vs_ssh_bulk}d) featuring hoppings $2J_1$ and $2J_2$. The spectra are plotted as a function of $J_2 / J_1$; here the ratio is manually tuned rather than derived from the driving parameters, in order to explore a broader parameter range. The bulk dispersion (shaded region), common to all panels, follows Eq.~\eqref{eq:bulk}, while different boundary terminations yield distinct energy spectra for the finite chains. The corresponding edge states are highlighted in a different color.

In the diamond chain, the doubling of the unit cell, now encompassing two plaquettes, introduces a boundary dependence on the parity of $N$. This is precisely reflected in the different edge-state behavior between panels~\ref{fig:diamond_vs_ssh_bulk}a (even $N$) and~\ref{fig:diamond_vs_ssh_bulk}c (odd $N$), especially near the gap-closing point $J_2 \sim J_1$. In the regime we are analyzing in this work ($\tilde{A} = 1.2)$, the ratio $J_2/J_1$ remains small (see Fig.\ref{fig:hopping_ren}b), and the spectra for even and odd $N$ become nearly identical.

The comparison with the SSH chain provides a clear interpretation of these features. In the SSH model, cutting the finite chain with an even number of sites (panel~\ref{fig:diamond_vs_ssh_bulk}b) or odd number (panel~\ref{fig:diamond_vs_ssh_bulk}d) link yields analogous differences in the behaviour of boundary states. Additionally, one may choose whether $J_1$ or $J_2$ terminates the boundaries for even $N$, which simply redefines the topological phase, shifting the presence of edge states to the regime $J_2/J_1 < 1$, but leaving the bulk spectrum unchanged.

\section{Symmetries of the driven chain \label{sec:symmetries}}

The undriven chain $\pi-$flux chain, described by the Hamiltonian in Eq. \eqref{eq:hamiltonian_or}, is known to preserve two nonsymmorphic symmetries \cite{Kremer2020}, namely, 
\begin{equation}
    \hat{\chi}(k) = \left( \begin{array}{ccc}
    1 & 0 & 0 \\
    0 & - e^{ik} & 0 \\
    0 & 0 & e^{ik}
    \end{array} \right),\quad \hat{\Pi}(k) = \left( \begin{array}{ccc}
    1 & 0 & 0 \\
    0 & e^{ik} & 0 \\
    0 & 0 & -e^{ik}
    \end{array} \right),
\end{equation}
for which the Hamiltonian satisfies:
\begin{eqnarray}
    \hat{\chi}(k)^{-1} \, \cdot \hat{H}_{\mathrm{rhomb}}(k) \cdot \chi(k) & = & - \hat{H}^*_{\mathrm{rhomb}}(k),\\
    \hat{\Pi}(k)^{-1} \, \cdot \hat{H}_{\mathrm{rhomb}}(k) \cdot \Pi(k) & = & \hat{H}^*_{\mathrm{rhomb}}(k),
\end{eqnarray}
up to gauge transformations. Identifying $\{ \mathcal{Z}_1,\mathcal{Z}_2,\mathcal{Z}_3 \}$ as the phase for each of the bands, labelled in increasing energy, one finds that the previous symmetries translate into the following constraints,
\begin{eqnarray}
    &\hat{\chi}(k):& \hspace{5pt} \mathcal{Z}_1 = \mathcal{Z}_3 \in\{0,\pm \pi/2\},\hspace{3pt} \text{and}\, \mathcal{Z}_2 \in \{0, \pi\},\\
    &\hat{\Pi}(k):& \hspace{5pt} \mathcal{Z}_1 + \mathcal{Z}_3 \in\{0, \pi\} ,\hspace{3pt} \text{and}\,\mathcal{Z}_2 \in \{ 0,\pi \}.
\end{eqnarray}

Interestingly, these symmetries are still preserved in the driven case as well, and can be applied to $\hat{H}_{\mathrm{eff}}(k)$ in Eq. \eqref{eq:h_eff_total}. First of all, let us recall that the hoppings in Eq. \eqref{eq:h_eff_total}, as shown in Appendix \ref{sec:hopping_ren}, can be replaced by $J_1 ,J_2 \rightarrow \vert J_1\vert,\vert J_2\vert$ after a gauge transformation. For real-valued hopping amplitudes, it is trivial to verify that 
\begin{eqnarray}
    \hat{\chi}(k)^{-1} \, \cdot \hat{H}_{\mathrm{eff}}(k) \cdot \hat{\chi}(k) & = & - \hat{H}^*_{\mathrm{eff}}(k),\\
    \hat{\Pi}(k)^{-1} \, \cdot \hat{H}_{\mathrm{eff}}(k) \cdot \hat{\Pi}(k) & = & \hat{H}^*_{\mathrm{eff}}(k).
\end{eqnarray}

As a function of $A$, the value of the Zak phases for each band remain pinned at $\{ \mathcal{Z}_1, \mathcal{Z}_2, \mathcal{Z}_3 \} = \{ \pi/2, \pi, \pi/2\}$.

\begin{figure}
    \centering
    \includegraphics{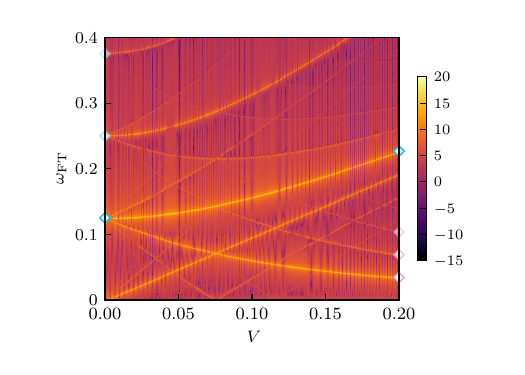}
    \caption{Fourier transform of entanglement entropy $S^{(A)}$ as a function of the interaction strength $V$. The color code corresponds to the amplitude of the corresponding oscillation frequency $\omega_{\text{FT}}$. The diamond markers correspond to the frequencies shown in Fig. \ref{fig:entanglement} with vertical lines, using the same color. Note that the harmonics of the active frequencies are highlighted as well, with a decreasing opacity. The parameters are the same as in Fig. \ref{fig:entanglement}.}
    \label{fig:fouriertransform_entanglement}
\end{figure}

\begin{figure}
    \centering
    \includegraphics{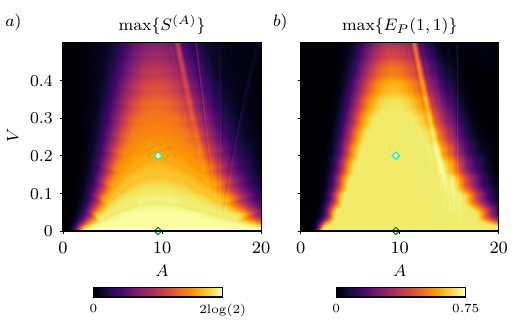}
    \caption{Maximum value of \textbf{a)} $S^{(A)}$ and \textbf{b)} $E_P(1,1)$ as a function of both $V$ and $A$, keeping the rest of the parameters as in Fig. \eqref{fig:entanglement} \label{fig:colormap_AV}. The diamond markers corresponds to the values of $(A,V)$ chosen for Fig. \eqref{fig:entanglement}.}
\end{figure} 

\section{Fourier transform for entanglement dynamics \label{sec:FT_entanglement}}

Figure~\ref{fig:fouriertransform_entanglement} shows the Fourier transform of $S^{(A)}$ as a function of $V$ and $\omega_{\mathrm{FT}}$, thereby providing a conceptual link to the behavior shown in Fig.~\ref{fig:entanglement}a) and c), which correspond to the values $V =0$ and $V = 0.2$, respectively. The diamond markers indicate the frequencies highlighted in Fig.~\ref{fig:entanglement}a) and c) by the vertical dashed lines, using the same color scheme. By progressively reducing the symbol opacity, we identify the corresponding higher harmonics of these frequencies in Fig.~\ref{fig:fouriertransform_entanglement}, as they appear in the Fourier spectrum. As the interaction is turned on, the single frequency governing the oscillations of $S^{(A)}$ 
 at $V = 0$ acquires additional contributions, giving rise to a more intricate dynamical behavior.\\

 On the other hand, Fig.~\ref{fig:colormap_AV} shows the maximum value of $S^{(A)}$ and $E_P(1,1)$ as a function of both $V$ and $A$, keeping the rest of the parameters as in Fig.~\ref{fig:entanglement}. The diamond markers correspond to the set of values $(V,A)$ chosen for Fig.~\ref{fig:entanglement}.



\bibliographystyle{apsrev4-2}
\bibliography{flo_deduped}
\end{document}